\documentclass[12pt]{iopart}

\usepackage{iopams}
\usepackage{amssymb}
\usepackage{amsfonts}
\usepackage{bm}
\usepackage{epsfig}
\usepackage{epsf}
\usepackage[]{graphicx}

\usepackage{color}

\usepackage{cite}
\usepackage{xspace}

\renewcommand{\d}{{\rm d}}

\newcommand {\bfb} {{\bf b}}

\newcommand {\bfr} {{\bf r}}

\newcommand {\bfv} {{\bf v}}

\newcommand {\bfB} {{\bf B}}
\newcommand {\bfE} {{\bf E}}
\newcommand {\bfF} {{\bf F}}

\newcommand {\bfbeta}  {\boldsymbol{ \beta}}
\newcommand {\bfnabla} {\boldsymbol{ \nabla}}

\newcommand {\E}  {{\varepsilon}}
\newcommand {\om}  {{\omega}}
\newcommand {\Om}  {{\Omega}}

\newcommand {\aTF}  {a_{\rm TF}}

\newcommand{\MBNExplorer}{\textsc{MBN Explorer}\xspace}

\begin{document}
\jl{2}


\title[Atomistic modelling with radiation reaction force included]
{Atomistic modelling of the channeling process with  
radiation reaction force included}

\author{Gennady B. Sushko, 
Andrei V. Korol
and 
Andrey V. Solov'yov
}

\address{
MBN Research Center, Altenh\"{o}ferallee 3, 60438 
Frankfurt am Main, Germany}

\begin{abstract}
Methodology is developed that incorporates
the radiative reaction force into the relativistic molecular dynamics
framework implemented in the \textsc{MBN Explorer} software
package.
The force leads to a gradual decrease in the projectile's energy $\E$
due to the radiation emission.
This effect is especially strong for
ultra-relativistic projectiles passing through oriented crystals where they
experience the action of strong electrostatic fields as has been shown in
recent experiments.
A case study has been carried out for the initial approbation of the
methodology developed.
Simulations of the processes of planar channeling and photon emission
have been performed for 150 GeV positrons in a 200 $\mu$m thick
single oriented Si(110) crystal.
Several regimes for the decrease in $\E$ have been established and
characterized.
Further steps in developing the code to include the necessary quantum
corrections are identified and possible algorithmic modifications are
proposed.
\end{abstract}

\section{Introduction \label{Introduction}}

In the last decade, a number of experiments have been carried out
at different accelerator facilities studying various phenomena related to
the interaction of ultra-relativistic beams of multi-GeV charged
projectiles with oriented crystals of different types and geometries
\cite{Lietti_EtAl-NIMB_283_p84_2014, Scandale_EtAl-PL_B719_p70_2013,
Bandiera_EtAl-PRL_v111_255502_2013, Bagli_EtAl-PRL_v110_175502_2013,
Bagli_EtAl-EPJC_v74_2740_2014, Bagli_EtAl-EPJC_v74_3114_2014,
Wienands_EtAl-PRL_v114_074801_2015, Wistisen_EtAl-PR-AB_v19_071001_2016,
Bagli_EtAl-EPJC_v77_71_2017, Wienands_EtAl-NIMB_v402_p11_2017,
Wistisen_EtAl-NatComm_v9_p1_2018, Scandale_EtAl-PR-AB_v21_014702_2018,
Scandale_EtAl-NIMB_v446_p15_2019, Scandale_EtAl-EPJC_v79_99_2019,
Wistisen_EtAl-PhysRevRes_v1_033014_2019,
Nielsen_EtAl-PhysRevD_v102_052004_2020}.
In these experiments the basic effect of channeling has been
exploited for the beam manipulation (beam steering, focusing,
collimation, splitting, and extraction) as well as for the
generation of intensive electromagnetic radiation.

The channeling phenomenon stays for a specific motion of
a charged projectile in an oriented crystal
along a crystallographic plane (planar channeling)
or an axis (axial channeling) due to correlated interactions with
the lattice atoms.
A comprehensive theoretical study by Lindhard \cite{Lindhard} has demonstrated that the penetration through a crystal strongly
depends on the orientation of the particle's velocity with respect to
crystallographic directions.
The continuum potential model \cite{Lindhard} explains the channeling
motion in terms of the action of a collective electrostatic field of the
atomic planes.\footnote{To be specific, we refer to the planar channeling.}
The field, being repulsive for positively charged particles, steers
them into the interplanar region so that they
move along the planar direction between two neighboring planes experiencing, simultaneously,
so-called channeling oscillations in the transverse direction.
Negatively charged particles, being attracted by the field,
channel in the vicinity of a plane.
Channeling motion may occur in linear crystals as well as in bent
\cite{Tsyganov1976} and periodically bent \cite{ChannelingBook2014} crystals.

Channeling oscillations give rise to a specific type of electromagnetic radiation, - the channeling radiation (ChR) \cite{ChRad:Kumakhov1976}.
A strongly correlated motion of the channeling particle in the collective
crystalline field results in a much higher intensity of ChR
as compared to the intensity of the incoherent bremsstrahlung
radiation emitted by the same projectile in the amorphous medium,
see e.g. \cite{Uggerhoj:RadEffDefSol_v25_p3_1993}.

A charged particle moving in a medium (more generally, in an external
field) loses its energy due to the emission of electromagnetic
radiation.
For ultra-relativistic electrons and positrons of very high energies
(tens of GeV and above) the radiative energy loss greatly
exceed the losses due to the ionizing collisions \cite{Landau4}.
Therefore, the radiation damping, i.e. the process of a gradual
decrease in the particle's energy due to radiation, must be accounted for an accurate quantitative analysis of the projectile motion.

Recent experiments with multi-GeV light projectiles (50 GeV positrons and 40-80 GeV electrons) passing through silicon and diamond oriented
crystals at small angles to the crystallographic axial and planar directions have shown the necessity to incorporate the radiative
recoil in describing the dynamics of the channeling particles
\cite{Wistisen_EtAl-NatComm_v9_p1_2018,Nielsen_EtAl-PhysRevD_v102_052004_2020}.
A theoretical scheme based on the continuous potential concept has been developed 
\cite{Wistisen_EtAl-PhysRevRes_v1_033014_2019,Nielsen_EtAl-PhysRevD_v102_052004_2020}
and convincing agreement with experiment has been demonstrated.

Various approximations have been used to simulate channeling phenomenon in oriented crystals.
Whilst most rigorous description relies on the framework of quantum mechanics (see Ref.
\cite{WistisenPiazza_PRD_v99_116010_2019} and references therein),
the classical description in terms of particles trajectories can be
applied to high-energy projectiles (hundred MeV range and above)
\cite{AndersenEtAl_KDanVidensk_v39_p1_1977,KKSG_simulation_straight}.
Simulation of the channeling and related phenomena within the framework
of continuum potential model has been implemented in several software
packages, see Ref. \cite{KorolSushkoSolovyov:EPJD_v75_p107_2021}
for a description of the most recent ones.
For the purpose of the current paper we mention the
code described in \cite{Dechan01} and a more recent one
presented in \cite{Nielsen_CPC_v252_107128_2020} (see also
\cite{Wistisen_EtAl-PhysRevRes_v1_033014_2019,
Nielsen_EtAl-PhysRevD_v102_052004_2020}).
These codes integrate the classical equations of motion of an
ultra-relativistic particle traversing oriented crystals with account
for (i) the continuous interplanar potential, and
(ii) the radiative damping force in the form presented in \cite{Landau2}.

Numerical simulations of the channeling and radiation processes beyond
the continuous potential framework can be carried out by means of the
multi-purpose computer package \textsc{MBN Explorer}
\cite{MBN_Explorer_2012}.
\textsc{MBN Explorer} is a software package for the advanced multiscale
simulations of structure and dynamics of complex molecular Meso-Bio-Nano
(MBN) systems.
It has many unique features and a wide range of applications in physics,
chemistry, biology, materials science, industry and medicine \cite{MBNExplorer_Book}.
It is suitable for classical non-relativistic and relativistic molecular
dynamics (MD), Euler dynamics, reactive \cite{MBNExplorer_ReactiveFF}
and irradiation-driven \cite{MBNExplorer_IDMD} MD simulations as well as
for stochastic dynamics or Monte Carlo simulations of various randomly
moving MBN systems or processes
\cite{MBNExplorer_KMC,Solovyov_EtAl-JCompChem_v43_p1442_2022}.
These algorithms are applicable to a large range of molecular systems of
different kind, such as nano- and biological systems, nanostructured
materials, composite/hybrid materials, gases, plasmas, liquids, solids,
and their interfaces, with the sizes ranging from atomic to mesoscopic.
The \textsc{MBN Explorer} package is supplemented with special multitask
software toolkit \textsc{MBN Studio} \cite{MBN_Studio_2019}  used to set up
and start MBN Explorer calculations, monitor their progress, examine calculation results, visualize inputs and outputs, and analyze specific characteristics determined by the output of simulations.

By means of a special module \cite{MBN_ChannelingPaper_2013} of the package
it is possible to simulate a passage of various particles (positively and
negatively charged, light and heavy) in various media, such as
hetero-crystalline structures (including superlattices), straight bent and
periodically bent crystals, amorphous solids, liquids, nanotubes,
fullerites, biological environment \cite{MBN_ChannelingPaper_2013}.
The applicability of the code to a particular structure can be adjusted
by choosing a proper interaction potential from a large variety of
the potentials already included in the package.
The software allows one to simulate the classical trajectories within the
framework of relativistic MD, i.e. when the speed of a projectile $v$ is
comparable to the speed of light, $c$.
Under such condition, in many cases the motion of even light particles
(electrons, positrons) can be treated in terms of classical mechanics
rather then the quantum mechanics.

Over the last years the module have been extensively applied to simulate
the propagation of ultra-relativistic charged particles in oriented
crystals accompanied by emission of intensive radiation.
A comprehensive description of the multiscale all-atom relativistic
molecular dynamics approach implemented in \MBNExplorer as well as a number
of case studies related to modeling channeling and photon emission by
ultra-relativistic projectiles  (within the sub-GeV up to ten GeV energy
range) in straight, bent and periodically bent crystals are presented
in a review article \cite{KorolSushkoSolovyov:EPJD_v75_p107_2021}.

Additional feature that has been implemented in \MBNExplorer and is discussed in Section \ref{Equations} below concerns the influence that have the radiation emitted by the particle upon its motion.
In terms of classical electrodynamics this influence can be described by
introducing a radiative reaction force acting on the projectile, see, e.g.
\cite{Jackson,Landau2}.
For ultra-relativistic electrons and positrons of very high energies
(tens of GeV and above) this force becomes quite noticeable so that it
must be accounted for to properly simulate the motion.
Several related case studies are presented in Section \ref{CaseStudies}.

\section{Methodology \label{Equations}}

The 3D simulation procedure of the motion an ultra-relativistic particle
in an external field or/and in an atomic environment is based on the
formalism of the classical relativistic mechanics and describes the motion in the laboratory reference frame.
\MBNExplorer implements the relativistic equations of motion written in the  following form:
\begin{eqnarray}
\left\{\begin{array}{l}
\displaystyle{
\dot{\bfv} =
{1 \over m \gamma}
\Bigl(\bfF -  \bfbeta\left(\bfF \cdot \bfbeta\right) \Bigr)}
\\
\dot{\bfr} = \bfv
\end{array}
\right. \ ,
\label{Equations:eq.01}
\end{eqnarray}
where $\gamma = \E/mc^2 =\left(1- v^2/c^2\right)^{-1/2}$
is the relativistic Lorentz factor of a projectile of energy $\E$ and mass $m$, $c$ is the speed of light, and $\bfbeta = \bfv/c$.

The force $\bfF$ acting on the projectile is the sum of two terms:
\begin{eqnarray}
\bfF = \bfF_{\rm em} + \bfF_{\rm rr} \,.
\label{Equations:eq.02}
\end{eqnarray}
The term $\bfF_{\rm em}$ stands for the total electromagnetic (Lorentz)
force due to (i) electrostatic field $\bfE$ created by atoms of the medium and/or by external sources of electric field, and (ii) external magnetic field $\bfB$:
\begin{eqnarray}
\bfF_{\rm em} = q\left( \bfE + \bfbeta\times \bfB\right) ,
\label{Equations:eq.03}
\end{eqnarray}
where $q$ is the charge of a projectile.

The second term on the right-hand side of Eq. (\ref{Equations:eq.02})
stands for the radiative reaction force, which can be
written in the following form \cite{Landau2}:
\begin{eqnarray}
\bfF_{\rm rr}
=
{2 q^2 \over 3 m c^3}
\Biggl\{
q\gamma
\left[
{\partial \bfE\over \partial t} + \left(\bfv\cdot\bfnabla\right)\bfE
+
\bfbeta\times
\left({\partial\bfB  \over \partial t} +  \left(\bfv\cdot\bfnabla\right)\bfB\right)
\right]
\nonumber\\
+
{q \over m c}
\Bigl[
\bfF\times \bfB
+
q \left(\bfbeta\cdot\bfE\right)\bfE
\Bigr]
-
{\gamma^2 \over m c} \,
\Bigl[
\bfF^2
-
q^2 \left(\bfbeta\cdot\bfE\right)^2
\Bigr]
\bfbeta
\Biggr\}
\label{Equations:eq.04}
\end{eqnarray}
where $\bfnabla$ stands for the vector differential operator
with respect to $\bfr$.

To accurately integrate highly non-linear equations of motion
(\ref{Equations:eq.01}) the fourth-order Runge-Kutta method with
adaptive time step control is implemented \MBNExplorer.

Applied to the propagation in an atomic environment the system  (\ref{Equations:eq.01} ) describes the classical motion of a particle in an electrostatic field due to the medium atoms.
In the absence of a magnetic field and for the stationary electric field
only the first term on the right-hand side of Eq. (\ref{Equations:eq.03}) is accounted for while the radiative reaction force reduces to
\begin{eqnarray}
\bfF_{\rm rr}
&=
{2 q^3 \over 3 m c^3}
\Biggl\{
\gamma \left(\bfv\cdot\bnabla\right) \bfE
+
{q \over m c}
\left(\bbeta\cdot\bfE\right)\bfE
-
{q \gamma^2 \over m c} \,
\Bigl[
\bfE^2
-
\left(\bbeta\cdot\bfE\right)^2
\Bigr]
\bbeta
\Biggr\}\,.
\label{Equations:eq.05}
\end{eqnarray}
Here $\bfE = \bfE(\bfr) = - q^{-1}\bnabla U({\bfr})$ with $U({\bfr})$
being the potential energy of the projectile's interaction
with the atoms:
\begin{eqnarray}
U(\bfr) =
\sum_{j} U_{\mathrm{at}}\left(\left|\bfr - \bfr_j\right|\right)
\label{Equations:eq.06}
\end{eqnarray}
where $U_{\mathrm{at}}$ are the potentials of individual atoms and
$\bfr_j$ denotes the position vector of the $j$th atom.
The code allows one to evaluate the atomic potential using the
approximations due to Moli\`{e}re \cite{Moliere} and Pacios
\cite{Pacios1993}.
A rapid decrease of these potentials with increasing the distances from the
atoms allows the sum~(\ref{Equations:eq.06}) to be truncated in practical
calculations.
The summation accounts only for the atoms located inside the sphere of a
(specified) cut-off radius $\rho_{\max}$ with the center at the instant
location of the projectile.
The value $\rho_{\max}$ is chosen large enough to ensure negligible
contribution to the sum from the atoms located at $r>\rho_{\max}$.
The search for such atoms is facilitated by using the linked cell algorithm
implemented in \MBNExplorer \cite{MBN_Explorer_2012,MBNExplorer_Book}.
More details on the algorithms implemented to compute trajectories of the
particles passing through a medium on the macroscopic scale are presented
in \cite{MBN_Explorer_2012,KorolSushkoSolovyov:EPJD_v75_p107_2021}.

The system (\ref{Equations:eq.01}) with $\bfF_{\rm em}=q\bfE(\bfr)$
and $\bfF_{\rm rr}$ from (\ref{Equations:eq.05}) can be applied to
any charged projectile moving in a specified atomic environment.
To be specific, in what follows when carrying out quantitative estimates
we consider the case of an
ultra-relativistic ($\gamma \gg 1$) positron ($q=e$) moving in
an oriented silicon crystal.

The applicability of (\ref{Equations:eq.05}) is subject to the condition
that the magnitude of the Lorentz force in the co-moving frame (in which
the particle is at rest) is small compared to $m^2c^2/e^2 = mc^2/r_0$
($r_0\approx2.818\times10^{-13}$ cm is the classical electron radius).
This condition leads to the following strong inequality that relates the electric field intensity $E$
and the particle's energy (see Eq. (76.5) in Ref. \cite{Landau2}):
\begin{eqnarray}
e E \ll (e E)_{\max}
= {1\over \gamma} { m c^2 \over r_0}
\approx {10^{6} \over \E}\,.
\label{Equations:eq.07}
\end{eqnarray}
On the right-hand side $\E$ is measured in GeV producing the value of
$(e E)_{\max}$ in GeV/cm.

For $\E=50$ and 150 GeV positrons  Eq. (\ref{Equations:eq.07}) produces
$(e E)_{\max}\approx 1.8\times10^4$ and $0.6\times10^4$ GeV/cm, respectively.
Describing the positron-atom interaction in terms of the screened Coulomb (Yukawa) potential
$Ze\exp(-r/\aTF)/r$ with atomic number $Z=14$ (a silicon atom)
and the Thomas-Fermi radius $\aTF=0.8853 Z^{-1/3}a_0 \approx 0.194$ \AA{},
one finds that the values of $(e E)_{\max}$ quoted above are achieved at
the distances $r \approx 0.063a_0$ and  $0.11a_0$, respectively.
In these estimates $a_0=0.529$  \AA{} is the Bohr radius.

General methodology implemented in \MBNExplorer to generate particles'
trajectories in a crystalline environment accounts for randomness in
sampling the incoming projectiles
(the key factors here are beam size, divergence and direction with respect
to the crystal orientation) as well as in displacement of the lattice atoms
from the nodal positions due to the thermal vibrations
\cite{MBN_ChannelingPaper_2013,
KorolSushkoSolovyov:EPJD_v75_p107_2021}.
As a result, each trajectory corresponds to a unique crystalline
environment and, therefore, all simulated trajectories are statistically independent and can be analyzed further to quantify the channeling process as well as the radiation emitted by projectiles.

The process of radiation emission is sensitive to the magnitude of
the quantum strong-field parameter $\chi$ defined as follows \cite{Landau4}
\begin{eqnarray}
\chi
=
\gamma {E \over E_0}
=
\gamma {eE \over F_0}
\label{Equations:eq.08} 
\end{eqnarray}
where $E_0 = 1.32\times10^{16}$ V/cm is the critical filed and
$F_0=eE_0=1.32\times10^{7}$ GeV/cm is the corresponding
critical force.
Classical description is valid if $\chi < 1$.
This implies the restriction on the range of distances between a
projectile and an atomic nucleus in individual close-encounter collisions.
Using, as above, the Yukawa potential to describe the positron--silicon atom
interaction one estimates the distances $r_{\rm min}$ beyond which the
classical framework is applicable:
\begin{eqnarray}
r > r_{\rm min}
\approx
\left\{
\begin{array}{ll}
0.29a_0 & \mbox{for $\E=50$ GeV}
\\
0.41a_0 & \mbox{for $\E=150$ GeV}
\end{array}
\right.
\label{Equations:eq.09}
\end{eqnarray}

\section{Case study \label{CaseStudies}}

In this section we present a case study that refers to
the $\E=150$ GeV positrons channeling in
an oriented Si(110) crystal (the interplanar distance $d=1.92$ \AA)
of thickness $L=200$ $\mu$m.
The positions of the crystal atoms were generated with account for
the atoms' random displacement from the nodes due to
thermal vibrations at temperature $T=300$ K.
For each atom the displacement was generated by means of the
normal distribution with the root-mean-square (rms) amplitude $u_T$
equal to 0.075 \AA{} \cite{Gemmel_RMP_v46_p129_1974}.
The direction of the incident particles, the  $z$-axis, is
chosen along the (110) plane and well away from
crystallographic axes of low indices to avoid the axial channeling regime.
A projectile enters the crystal at $z=0$ and exits at $z=L$.
The crystal is considered infinitely large in transverse $x$ and $y$
directions.
The $y$-axis is aligned with  $\langle 110 \rangle$ axial direction.
At the entrance the $x$ and $y$ coordinates of the particle
have been randomly generated by means of the uniform distribution within
the rectangle $\Delta y = 4d= 7.68$ \AA{}, $\Delta x = 2a= 10.86$ \AA{}
where $a=5.43$ \AA{}  is the lattice constant.
The angular distribution of the positrons at the crystal entrance has been
ignored, therefore, the results presented below refer to an ideally
collimated beam of zero divergence.

The trajectories have been obtained by integrating
classical equations of motion (\ref{Equations:eq.01}).
Therefore, it was expected that in the course of molecular dynamics
simulation of the particles' propagation through the
crystalline medium events of the close-encounter collisions with the
crystal atoms would occur in which either condition (\ref{Equations:eq.07})
or condition (\ref{Equations:eq.09}) will not be met.
Such events lead to a (noticeable) overestimation of the energy
loss due to the radiation emission in a localized spatial region.
Hence, important was to evaluate the fraction of the trajectories affected
by such collisions.

\begin{figure}[ht]
\centering      
\includegraphics[scale=0.5,clip]{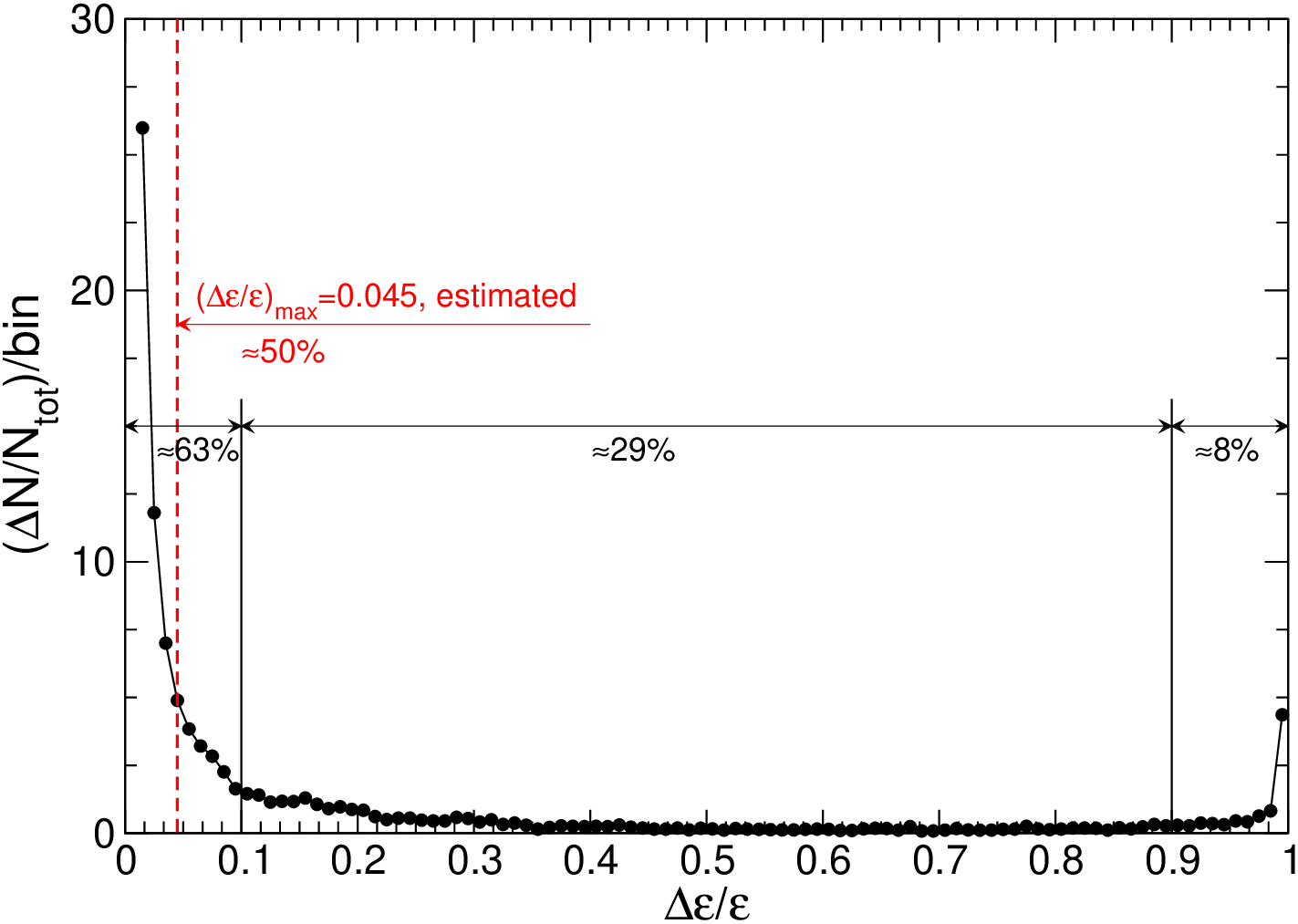}
\caption{
Distribution (a histogram) of positrons at the crystal exit with respect to
the relative energy loss, $\Delta \E/\E$.
Symbols indicate the heights of the histogram rectangles placed in the center
 of the bins (the bin width is 0.01).
 Solid vertical lines mark three intervals of $\Delta \E/\E$.
 The values indicated in per cent correspond to the fractions
 of the total number of trajectories, $N_{\rm tot}$, moving along which
a particle experiences the energy loss within each interval.
 The vertical dashed line marks the maximum energy loss
 $(\Delta \E/\E)_{\max}$
 estimated within the continuous potential framework.
 Approximately 50 per cent of all trajectories fall in the
interval $[0, 0.045]$.
}
\label{Figure01.fig}
\end{figure}

Figure \ref{Figure01.fig} shows a distribution
$\Delta N/N_{\rm tot}/bin$ of the particles leaving the crystals
as a function of a relative energy loss $\Delta \E/\E$.
Here $N_{\rm tot}\approx 10^4$ is the total number of the
simulated trajectories, $\Delta N$ is the number of trajectories that
correspond to a particular interval of the energy loss; the width of
each interval (bin) is 0.01.
Symbols show the heights of the histogram rectangles (not drawn) and are
placed in the bins' centers.
Solid vertical lines divide the whole interval $\Delta \E/\E=[0,1]$ into
three subintervals and the numbers with  per cent sign show the fraction of
$N_{\rm tot}$ that falls into each subinterval.
Dashed (red) vertical line marks the maximum relative loss equal to 0.045
that can be estimated analytically applying the continuous potential model
to describe the positron channeling (see Section \ref{Estimation} for the
details).

The distribution presented shows that most of the particles, 63 per cent,
experience the energy loss $\Delta \E/\E \leq 0.1$ that is approximately
in accordance with the estimate.
However, a noticeable number of the particles, ca 20 \%, lose
anomalously large fraction of the initial energy ($\Delta \E/\E > 0.3$)
due to the action of the radiative reaction force.
Hence, it is informative to carry out more detailed analysis of the
simulated trajectories aiming at revealing the reason for large values of
the losses.

\begin{figure}[ht]
\centering
\includegraphics[scale=0.52,clip]{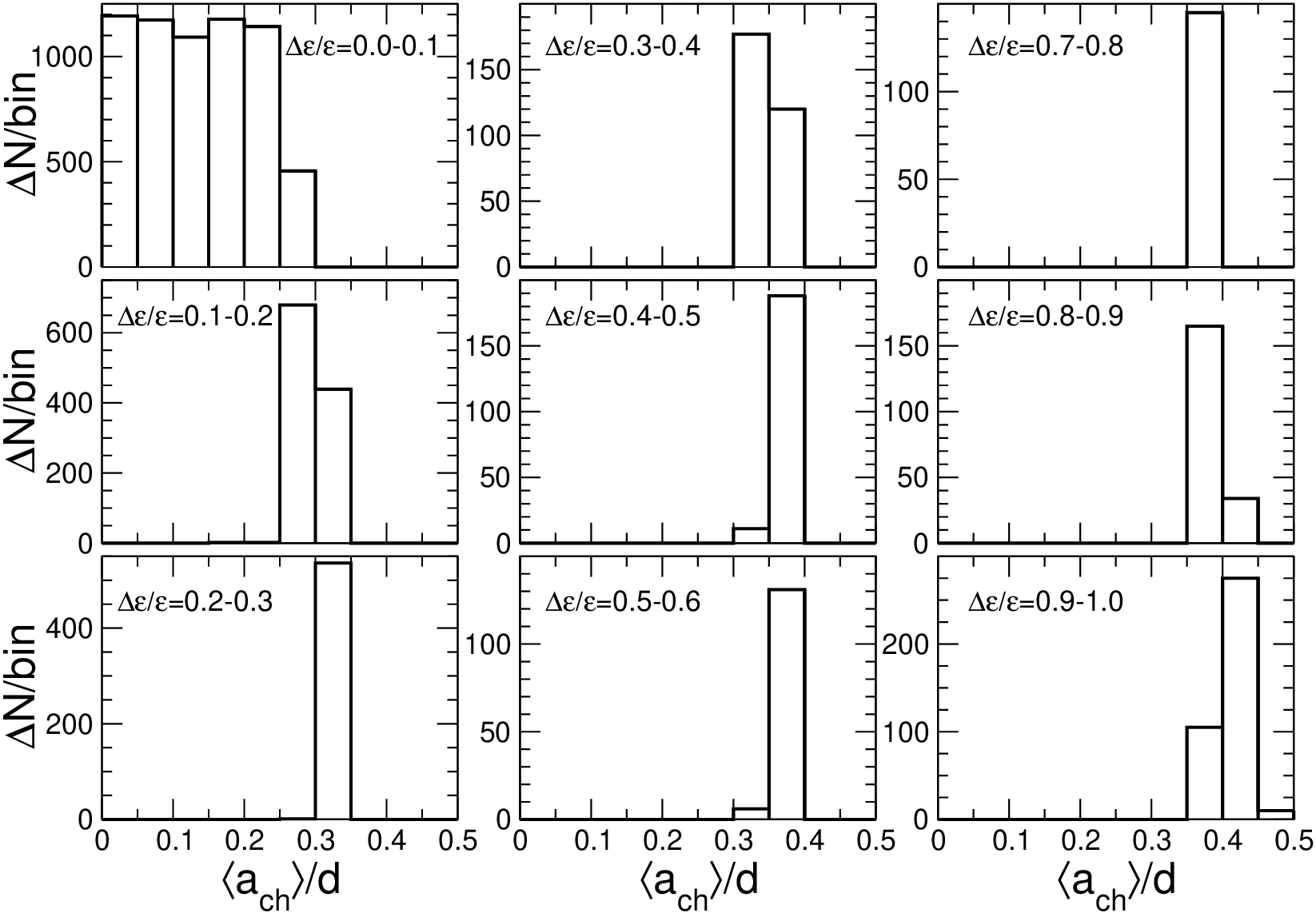}
\caption{
Distribution of projectiles with respect to the average amplitude of
channeling oscillations, $\langle a_{\rm ch} \rangle$,
along the trajectory.
Nine graphs correspond to different intervals of $\Delta \E/\E$ as indicated in
each graph.
}
\label{Figure02.fig}
\end{figure}

Figure \ref{Figure02.fig} presents the distribution
of the number $\Delta N$ of particles, which channel through the whole
crystal, with respect to the average amplitude $\langle a_{\rm ch} \rangle$
of channeling oscillations calculated along the particles' trajectories.
The average amplitude is measured in units of the interplanar distance $d$
so that $\langle a_{\rm ch}\rangle_{\max}/d=0.5$.
Each graph corresponds to the indicated interval of
$\Delta \E/\E$.
Positron channeling occurs between two neighbouring atomic planes.
Hence, the larger channeling amplitude is the closer a projectile
approaches the planes experiencing the action of stronger field in this
spatial domain.
However, significant energy losses happen not due to the action of the
average electrostatic field of a plane but rather during rare individual
collisions of the projectile with the atoms that are displaced from the
nodal positions due to the thermal vibrations.
In more detail this issue is addressed below in the paper, see discussion
in connection with Figs. \ref{Figure04.fig} and
\ref{Figure05.fig} below.

\begin{figure}[ht]
\centering
\includegraphics[scale=0.52,clip]{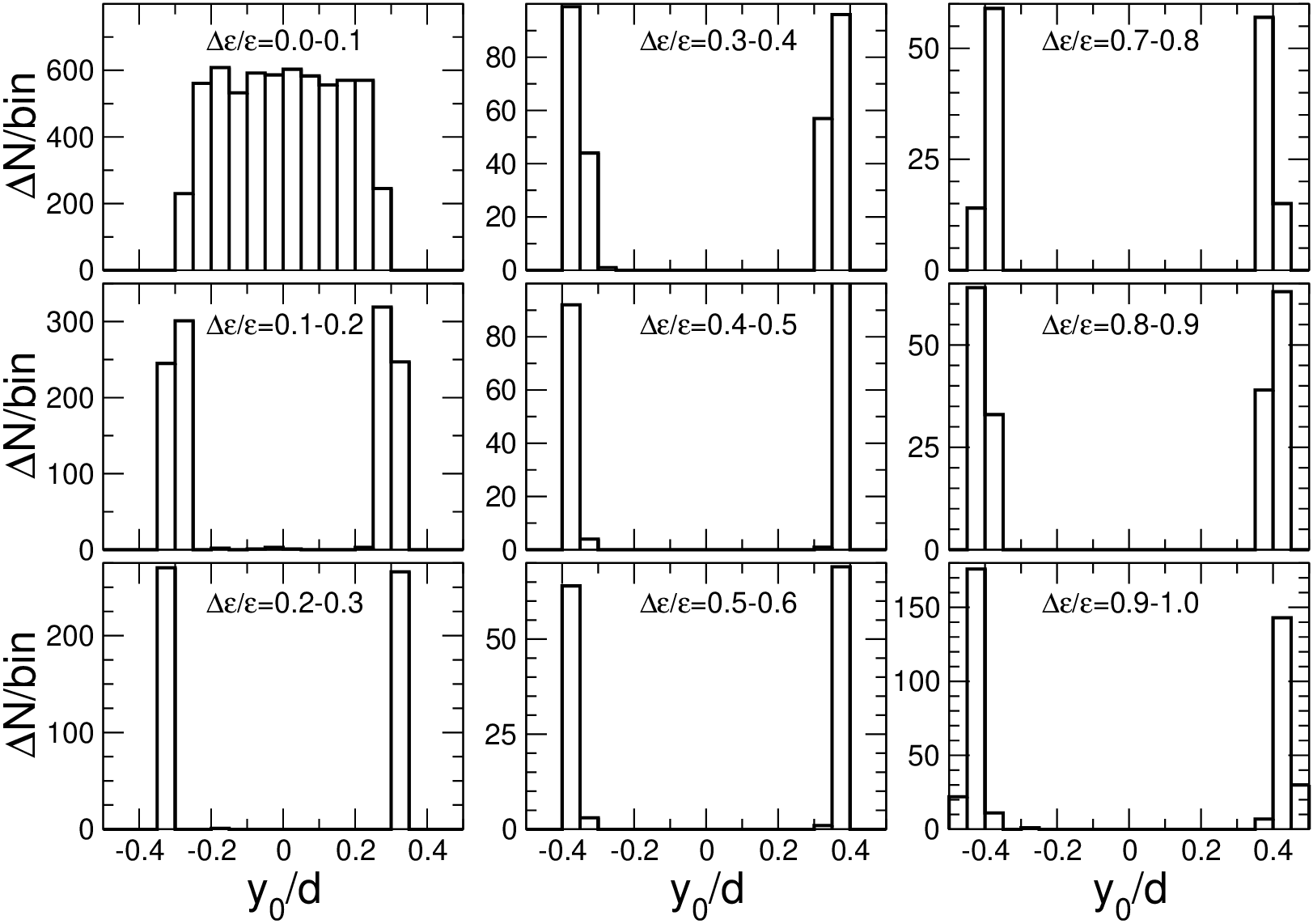}
\caption{
Distribution of projectiles with respect to the coordinate $y_0$ at the
crystal entrance.
Nine graphs correspond to different intervals of $\Delta \E/\E$ as
indicated in each graph.
The distributions presented correlate with those shown in Fig.
\ref{Figure02.fig}.
}
\label{Figure03.fig}
\end{figure}

 The dechanneling length $L_{\rm d}$ of a 150 GeV positron in Si(110),
 estimated using Eq. (1.50) from Ref. \cite{BiryukovChesnokovKotovBook},
 is equal to 8.2 cm, i.e. two orders of magnitude larger than the crystal
 thickness $L$ used in the simulations.
 Therefore, most of the particles accepted into the channeling mode at the
 entrance channel through the whole crystal and their
channeling amplitude does not change noticeable
 in the course of propagation through the crystal.
Therefore, for a particle of an ideally collimated beam, the average
amplitude is  equal, approximately, to the absolute value of the transverse
coordinate $y_0$  measured at the entrance with respect to the geometrical
center of the  channel,  $\langle a_{\rm ch}\rangle \approx |y_0|$.
Figure \ref{Figure03.fig} presents the distribution of the channeling
particles with respect to the transverse coordinate $y_0$ (measured in
units of the interplanar distance).
Comparing the graphs that refer to the same value of  $\Delta \E/\E$
in Figs. \ref{Figure02.fig} and \ref{Figure03.fig} one notices a strong
correlation between the distributions.
To be noted, in particular, are relatively small numbers of particles that
enter the crystal with $|y_0|/d \geq 0.45$ or/and move in the channeling
mode with $\langle a_{\rm ch}\rangle_{\max}/d \geq 0.45$, see the
right-bottom graphs in both figures.
This range of the coordinates / amplitudes corresponds to the distances
from the crystallographic plane approximately equal to the rms amplitude
$u_T=0.075$ \AA.
In this domain the volume density of the crystal atoms is enhanced due to
thermal vibrations so that there is high probability for a projectile to
experience a hard collision with the nuclei and, as a result, to leave the
channeling mode.

\begin{figure}[ht]
\centering
\includegraphics[scale=0.52,clip]{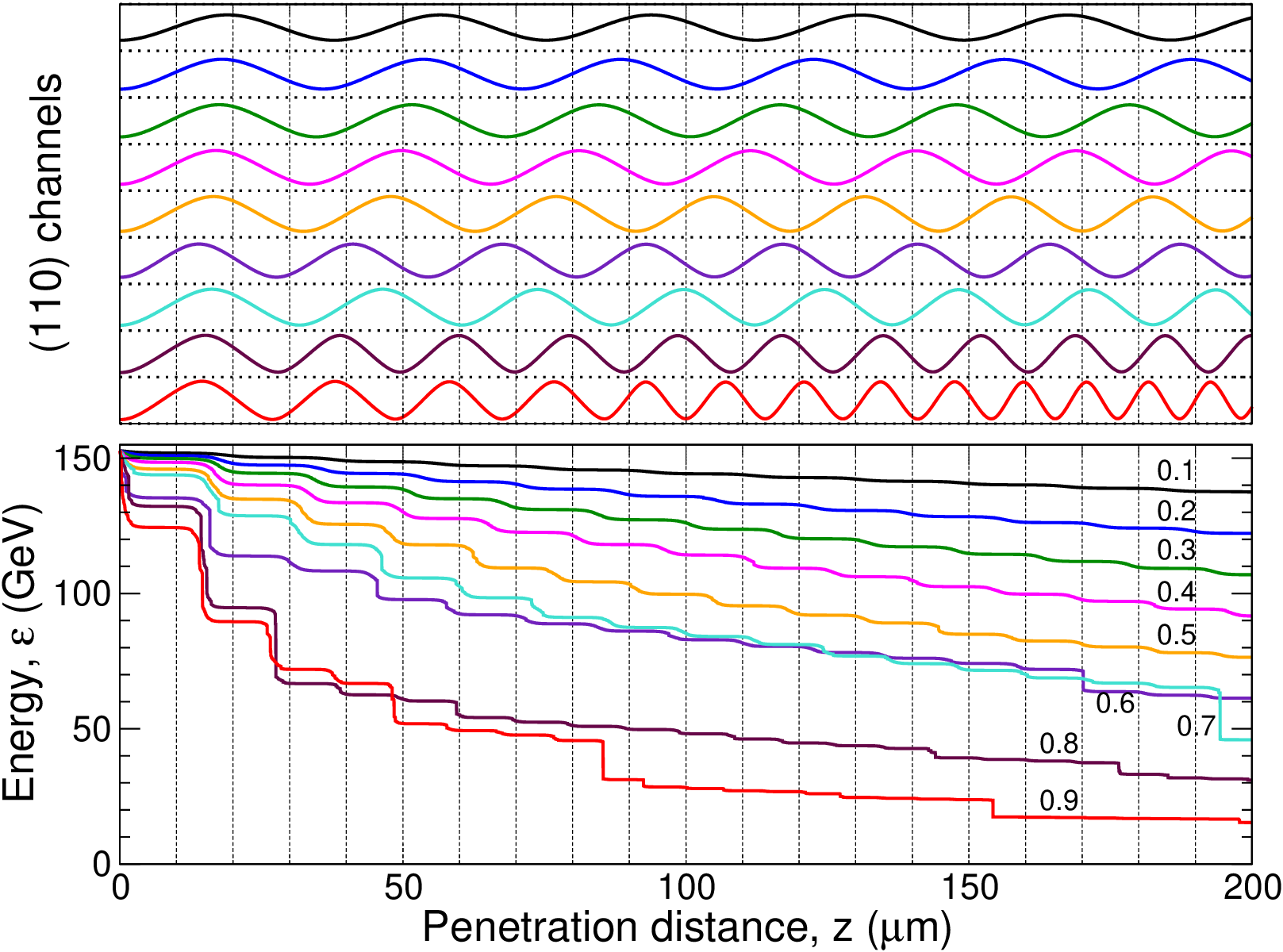}
\caption{
\textit{Top:}  Selected simulated trajectories (in projection on the
$(yz)$-plane) of 150 GeV positrons in an oriented silicon(110) crystal.
The $z$-axis is directed along the incoming projectiles,
the $(xz)$-plane is parallel to the crystallographic planes
(dotted horizontal lines), and the $y$-axis is perpendicular to the planes.
The trajectories correspond to different values of the relative energy
loss: Starting with $\Delta \E/\E=0.1$ for the top trajectory up to
$\Delta \E/\E=0.9$ for the lowest trajectory with the increment 0.1.\\
\textit{Bottom:} Dependencies $\E=\E(z)$ of the projectile energy
on the penetration distance $z$ calculated for the trajectories shown on
the top graph.
The value of $\Delta \E/\E$ is indicated for each dependence.
}
\label{Figure04.fig}
\end{figure}

Within the framework of molecular dynamics, the simulation of a
projectile's motion is based on solving the equations of motion
(\ref{Equations:eq.01}) accounting, as in reality, for the interaction
of the projectile with individual atoms of the crystal.
The potential energy of this interaction varies rapidly in the course of
the projectile's motion in the vicinity of the atomic chains and so does
the radiative reaction force $\bfF_{\rm rr}$ (\ref{Equations:eq.05}).
Analysis of each simulated trajectory allows one to establish the
spatial regions where the action of $\bfF_{\rm rr}$ results in significant
energy losses.

The top graph in Fig. \ref{Figure04.fig} presents several simulated
trajectories (in their projections on the $(yz)$ plane) that correspond to
different values of the relative energy loss starting with
$\Delta \E/\E=0.1$ for the uppermost trajectory up to  $\Delta \E/\E=0.9$
(the lowest trajectory) with the increment 0.1.
For the sake of convenience the trajectories are shown in separate
channels, the geometric borders of which are drawn with dotted lines.
The feature to be noted is the decrease in the period $\lambda_{\rm ch}$ of
channeling oscillations with penetration distance $z$.
The dependence $\lambda_{\rm ch} \propto \sqrt{\E(z)}$ is most clearly
seen when comparing the periods in the initial and final segments of the
trajectories with $\Delta \E/\E\geq 0.5$ and relates the segments to the
corresponding parts of dependencies $\E=\E(z)$ that are shown in the
bottom graph.
The value of $\Delta \E/\E$ is indicated for each dependence.
Vertical grid lines, drawn in each graph, help one to correlate the
$\E=\E(z)$ dependencies with the trajectories.

Two upper trajectories that correspond to $\Delta \E/\E=0.1$ and 0.2
refer to the channeling oscillations with the average amplitudes
$\langle a_{\rm ch}\rangle/d$=0.25 and 0.32, respectively.
Hence, they stay away from the atomic planes and experience, on average,
the action of a relatively uniform interplanar field which results, in
turn, in a smooth gradual decrease of the energy with the penetration
distance.

Larger values of the relative energy loss correspond to the motion with
larger amplitudes, see Fig. \ref{Figure02.fig}.
Projectiles that move along such trajectories spend more time in close
vicinity to the atomic plane and thus have higher probability to
successively collide with several atoms.
In each collision the particle experiences a localized action of
much stronger atomic field.
As a result, the dependence $\E=\E(z)$, being still a monotonously
decreasing one, becomes modulated: the segments of a relatively weak
decrease, which corresponds to the motion well away from the planes,
alternate with the segments of a steeper decrease due to the collisions
with atoms.
The modulations of this type  are well pronounced in the  $\E=\E(z)$
dependencies shown in Fig. \ref{Figure04.fig}\,\textit{bottom}
for $\Delta \E/\E=0.3$, 0.4, and 0.5.
The period of modulation equals to the half-period of the channeling
oscillations, see the corresponding trajectories in the top graph.
The length of the segments of the steeper decrease is of the order of
several microns, which means that the projectile experiences a number
of sequential collisions.

\begin{figure}[ht]
\centering
\includegraphics[scale=0.52,clip]{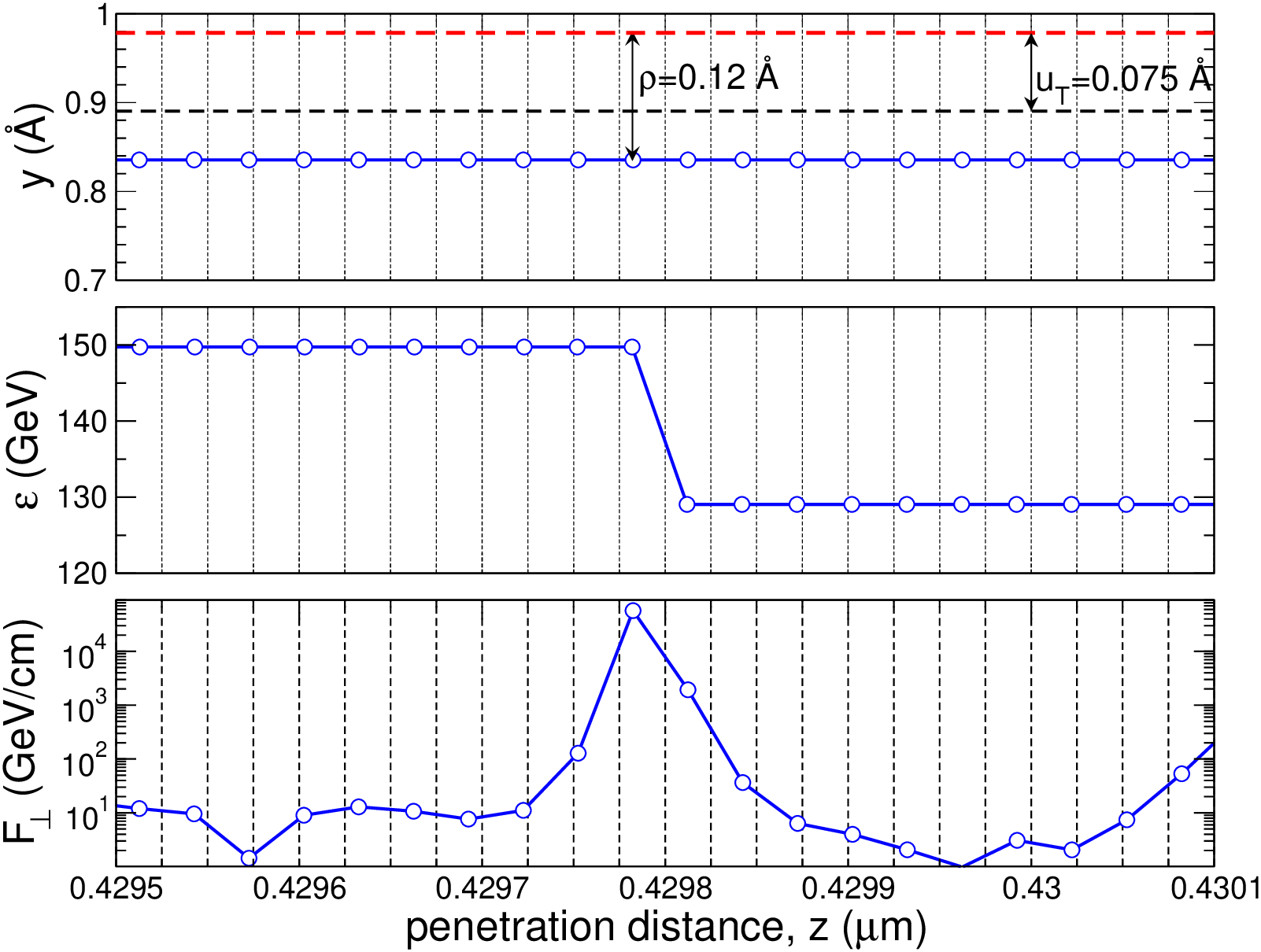}
\caption{
\textit{Top:}  Solid line with open circles shows a short segment of the
simulated trajectory of a 150 GeV positron in the vicinity,
$\rho=0.12$ \AA{}, of a geometric border (thick dashed line) of the Si(110)
planar channel.
Thin solid line marks the distance from the border equal to the rms thermal
vibration amplitude $u_T$.
The transverse coordinate $y$ is measured from the channel geometric
center.\\
\textit{Middle:} The projectile's energy versus the penetration distance,
$\E=\E(z)$.
\\
\textit{Bottom:} The transverse component of the electrostatic force acting
on the particle.
}
\label{Figure05.fig}
\end{figure}

Significant energy loss can occur in a close collision of a projectile
with a constituent atom.
Such events reveal themselves as sharp step-like change of the function
$\E=\E(z)$, see the curves for $\Delta \E/\E \geq 0.6$ in Fig.
\ref{Figure04.fig}\,\textit{bottom}.
In more detail this feature is illustrated by Fig. \ref{Figure05.fig},
which shows the segment of the simulated trajectory (top), dependence
$\E(z)$ (middle) and the transverse component of the total electrostatic
force acting on the projectile (bottom) over a short path, $\Delta z=6$
\AA{}.
In the course of simulations the equations of motion
(\ref{Equations:eq.01}) were integrated with the time step
$\delta t=10^{-4}$ femtoseconds, which was also set as the output step.
The open circles in the figure represent the output data that are separated
by $\delta z \approx c\delta t \approx 0.3$ \AA.

The dependence $\E=\E(z)$ exhibits a sharp decrease,
$\Delta \E/\E \approx 0.2$, at $z\approx 0.42978$ $\mu$m due to a single
close collision with a crystal atom.
In such a collision the impact parameter $b$ is smaller than the atomic
screening radius, therefore, the collision effectively occurs in the point
Coulomb field of the nucleus.
Using Eqs. (\ref{Yukawa:eq.06}) and (\ref{Yukawa:eq.07}) one estimates
(within the classical framework) the impact parameter that corresponds to
the quoted value of the relative energy loss:
$b\approx 0.09\aTF \approx 1.7\times 10^{-2}$ \AA{}.\footnote{The
scattering angle in such collision, calculated using Eq.
(\ref{Classical:eq.02}), $\theta_s \approx 0.16$ $\mu$rad is small compared
to the Lindhard critical angle $\theta_L =\sqrt{2U_0/\E}\approx 17$
$\mu$rad, where $U_0\approx 22$ eV is the depth of the continuous potential
well in Si(110), see Table \ref{Table.3}.}
Moving along the the trajectory shown in Fig.
\ref{Figure04.fig}\,\textit{top} the projectile passes at
the distance $\rho=0.12$ \AA{} from the geometric border of the channel.
This value is order of magnitude larger than the impact parameter,
therefore, to make the collision happen the atom must be displaced from its
nodal position by distances within the interval $[\rho-b, \rho+b]$.
The probability $\Delta P(\rho,b)$ of such displacement one estimates
by means of the Gaussian distribution:
$\Delta P(\rho,b) \approx (2b /\sqrt{2\pi u_T^2})
\exp\left(-{\rho^2 / 2u_{\rm T}^2}\right) \approx
5\times 10^{-2}$.
This estimate shows, that the probability of a large energy loss in a
single collision is high enough for projectiles experience channeling
oscillations with sufficiently large amplitude,
$a_{\rm ch}\gtrsim d/2- ku_T$ with $k=1\dots 3$.
A particle that channels with large amplitude over a macroscopic distance
can experience several events of close collisions, see the dependencies
corresponding to $\Delta \E/\E \geq 0.6$ in Fig. \ref{Figure04.fig}.

Figure \ref{Figure05.fig}\,\textit{bottom} shows the variation of the
transverse component $F_{\perp}$ of the electrostatic force.
As the particle approaches the atom the force increases sharply reaching
the value $F_{\perp}\approx 7\times 10^4$ GeV/cm in the point of the
collision.
Such a force does not satisfy the condition (\ref{Equations:eq.07})
and also it results in a large value $\chi\gg 1$ of the quantum
strong-field parameter, Eq. (\ref{Equations:eq.08}).
Therefore, in the vicinity of the points of close collisions the classical
approach to the particle dynamics and the radiation emission is not
applicable.
Instead, the description within the framework of quantum electrodynamics
must be used.
However, important is that the quantum collisional and radiation processes
are random, fast and localized in space.
Therefore, they can be incorporated into the framework of classical
molecular dynamics in a stochastic manner with the probabilities elaborated
on the basis of quantum mechanics / electrodynamics.
This can be achieved because the aforementioned quantum processes happen on
the sub-femtosecond time scales (i.e., during the periods comparable or
smaller than a typical single time step of the simulations).
Such methodology has been already implemented in the \MBNExplorer software
package for atomistic simulations of the irradiation-driven transformations
of complex molecular systems \cite{MBNExplorer_IDMD}.

\begin{figure}[ht]
\centering
\includegraphics[scale=0.52,clip]{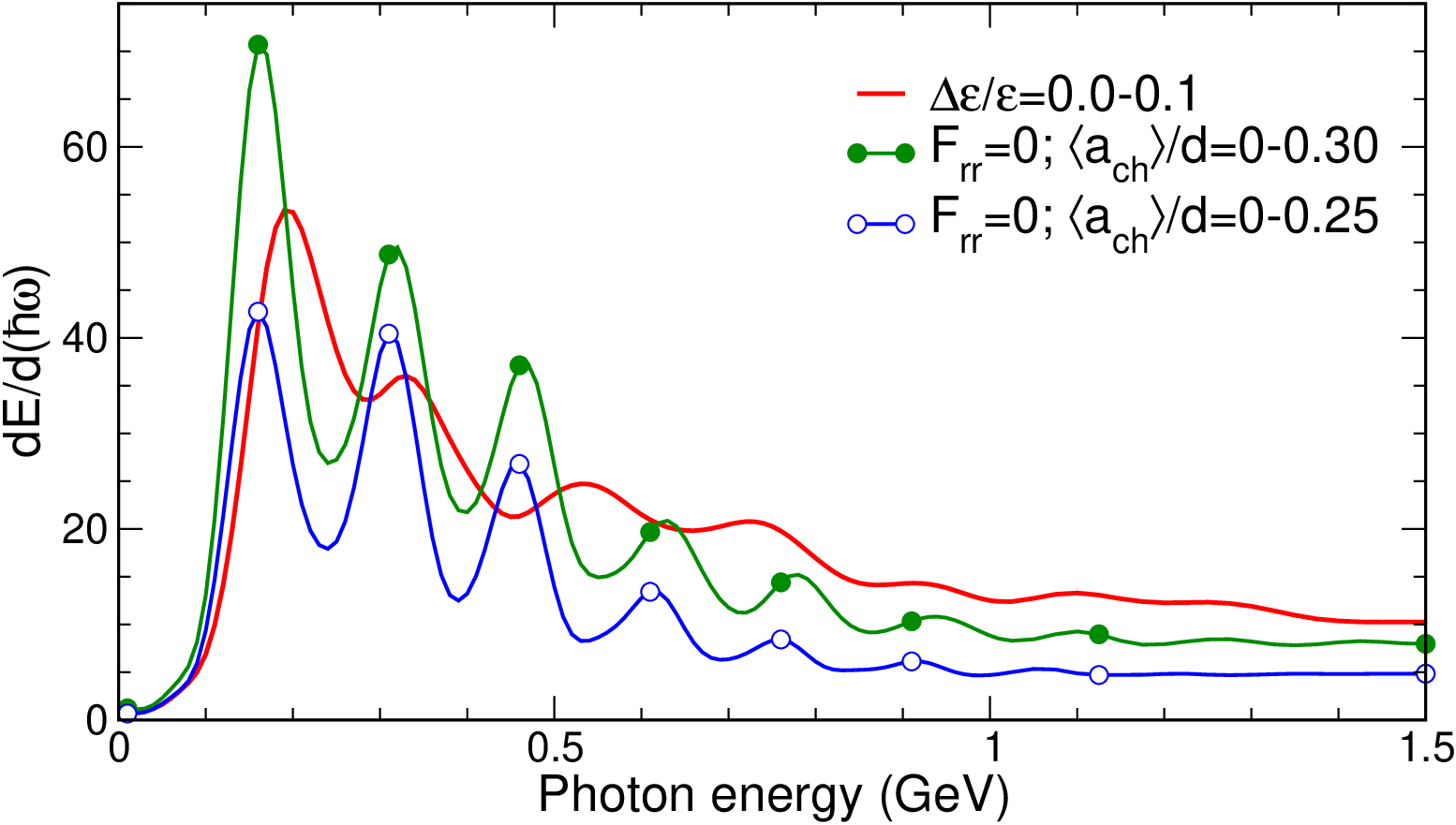}
\caption{Spectral distribution of channeling radiation emitted by 150
GeV positrons in 200 $\mu$m thick Si(110) crystal.
Solid (red) curve without symbols presents the dependence calculated
for the trajectories that correspond to the interval
$\Delta \E/\E=0-0.1$ of the relative energy losses due to the
radiative reaction force $F_{\rm rr}$.
Solid curves with symbols show the spectra calculated for the trajectories
without account for $F_{\rm rr}$ and selected with respect to the ranges
for the average amplitude of channeling oscillations, $\langle a_{\rm ch}
\rangle$, as indicated in the legend.
See also explanations in the text.
}
\label{Figure06.fig}
\end{figure}

Most of the simulated trajectories are free from the aforementioned
difficulties and can be used for the statistical analysis of other
processes, for example, the radiation emission.
Figure \ref{Figure06.fig} presents spectral distribution
$\d E/\d(\hbar\om)$ of the channeling radiation emitted by 150 GeV
positrons within the cone $\theta=6/\gamma \approx 20$ $\mu$rad
along the direction of incident particles.
The calculations have been performed for the trajectories simulated
with (solid curve without symbols) and without (solid curves with
symbols) account for the radiative reaction force $F_{\rm rr}$.
In the former case only those trajectories were taken into account that
satisfied the condition $\Delta \E/\E \leq 0.1$.
Top left graph in Fig. \ref{Figure02.fig} indicates that the condition
is fulfilled for all trajectories for which
average amplitude of the channeling oscillations does not exceed
$d/4$ as well as a large fraction of the trajectories for which
$0.25 < \langle a_{\rm ch}\rangle /d  \leq 0.3$.
Overall number of such trajectories is nearly two thirds of the
total number of simulated trajectories, see
Fig. \ref{Figure01.fig}.
The resulting spectral dependence was calculated as the average
of the spectra emitted from each trajectory.
The details of the numerical procedures can be found in Refs.
\cite{MBN_ChannelingPaper_2013,KorolSushkoSolovyov:EPJD_v75_p107_2021}.

To assess the extent to which the radiative reaction force affects the
radiation emission one has to simulate trajectories assuming
$F_{\rm rr}=0$ and calculate the resulting spectral distribution.
However, prior to calculating the distribution a certain procedure
must be applied to select the trajectories that can be
matched to those accounted for when calculating
$\d E/\d(\hbar\om)$ with $F_{\rm rr}\neq 0$.
The criterion for the selection relies on matching the
average amplitude of the channeling oscillations.
Two curves with symbols in Fig. \ref{Figure06.fig} show the spectral
distributions calculated by averaging the distributions of radiation
emitted from the trajectories for which the value of
$\langle a_{\rm ch}\rangle/d$ falls into the ranges indicated in
the caption.
All three curves exhibit the undulator-type spectral dependence:
the presence of equidistant peaks that correspond to the emission
into several first harmonics of the channeling radiation.
In the absence of the radiative reaction force the channeling oscillations
of the positrons are nearly harmonic, so that their frequency
$\Om_{\rm ch}$ does not depend on the amplitude $a_{\rm ch}$ and stays
constant in the course of motion.
This results in a series of well-pronounced and well separated peaks the
positions of which $\om_{n}$ ($n=1,2,3,\dots$) are related to
$\Om_{\rm ch}$ through $\om_n = 2\gamma^2 n\Om_{\rm ch}/(1 + K^2/2)$,
where $K\approx2\pi \gamma a_{\rm ch}/\lambda_{\rm ch}$ is the
so-called undulator parameter, $\lambda_{\rm ch}$ stands for the spatial
period of the oscillations.
The radiative reaction force, acting mainly in the longitudinal direction,  decelerates the particle leading to the increase in the channeling oscillations frequency along the trajectory,
$\Om_{\rm ch}(z) \propto \Bigl(\E(z)\Bigr)^{-1/2}$.
This results in the peaks broadening and shift towards higher energies.

\section{Conclusion and outlook\label{Conclusion}}

We have presented the algorithm developed to incorporate the radiative
reaction force into the relativistic molecular dynamics framework.
The methodology relies on integrating the equations of motion of an
ultra-relativistic particle moving in an external electromagnetic field
or passing through a medium.
In the latter case, the particle's dynamics is subject to the interaction
with all atoms of the medium.
The action of the radiative reaction force leads to a gradual decrease in
the particle's energy in the course of its motion due to the emission of
radiation.
The effect of the radiative recoil becomes especially significant for
ultra-relativistic projectiles passing through oriented crystals where they are subject to the strong electrostatic crystalline fields.
Recent channeling experiments with tens of GeV electrons and positrons
\cite{Wistisen_EtAl-NatComm_v9_p1_2018,
Wistisen_EtAl-PhysRevRes_v1_033014_2019,
Nielsen_EtAl-PhysRevD_v102_052004_2020}
have shown the necessity to account for the radiative reaction force in
describing the dynamics of the particles.

A case study has been carried out for the initial approbation of the
methodology and its implementation in the \textsc{MBN Explorer} software
package.
The simulation of the planar channeling process along with the calculation
of spectral distribution of the emitted radiation have been performed for
150 GeV positrons passing through a 200 $\mu$m thick single oriented
Si(110) crystal.
Special attention has been devoted to the trajectory-by-trajectory analysis
of the variation of the particle's energy $\E$ with the penetration distance
$z$.
Several regimes for the decrease of $\E(z)$ have been established.
One of these corresponds to the channeling motion well away from the
atomic planes that are diffused due to thermal vibrations of the atoms.
This regime accommodates two thirds of the channeling particles.
The radiative energy loss, calculated for the simulated trajectories, is in
accordance with the estimate derived using the continuum potential model.
The trajectories that have segments close to the diffused atomic planes
reveal steeper decrease in $\E(z)$ due to the events when several sequential
distant collisions with the atoms occur along the path.
Finally, in a limited number of the trajectories the events of close
collisions with the atomic nuclei have been identified.
In these collisions the impact parameter (i) is small enough to allow for a
noticeable decrease  in $\E(z)$ over the sub-angstrom path, but (ii) large
enough for the scattering angle to be much less than the Lindhard critical
angle.
Hence, after the collision the particle still moves in the channeling mode but
with lower total energy.

In the spatial regions where a projectile experiences aforementioned
close-encounter collisions the conditions  (\ref{Equations:eq.07}) and/or
(\ref{Equations:eq.09}) are not met so that the classical description
of the radiative energy losses in terms of the radiative reaction force
(\ref{Equations:eq.05}) as well as the description of the particle's motion in
terms of trajectories  become inadequate.
In these regions rigorous treatment can
only be achieved by means of quantum mechanics.
However, taking into account that such events are random, fast and local
they can be incorporated into the classical molecular dynamics framework
according to their probabilities that are related to the cross sections
(i) of the elastic scattering of the projectile from an isolated atom, and
(ii) of the bremsstrahlung emission in the static atomic field, see, e.g.
\cite{Landau4}.
Similar methodology has been already implemented in \MBNExplorer within the
framework of the irradiation-driven molecular dynamics
\cite{MBNExplorer_IDMD}.\footnote{To be noted that account for random
events of inelastic scattering of a projectile from individual atoms
that lead to atomic excitation / ionization and to a random change in
the direction of the particle's velocity has been already
implemented in \textsc{MBN Explorer} following the algorithm described some
time ago in Ref. \cite{Dechan01}.}
Algorithmically, to identify these regions, at each step of integration of the
equation of motion  (\ref{Equations:eq.01}) the validity of inequalities
(\ref{Equations:eq.07}) and $\chi < 1$ (see (\ref{Equations:eq.08})) must be
checked.
If these conditions are not met then the deterministic classical description
is substituted, locally, with the random generation of the
kinematic and dynamic quantities in accordance with the known
probabilities of the aforementioned processes.
We plan to introduce these new features to the numerical algorithm
aiming to expand the range of applicability of the code for the
trajectories simulations as well as for the calculation of the spectra.
Another feature that will be implemented concerns the calculations of the
multiphoton emission.
This process becomes important for high-energy projectiles channeling in
very thick (up to the centimeter range) crystals
\cite{Nielsen_EtAl-PhysRevD_v102_052004_2020}.

\ack

The work was supported by Deutsche Forschungsgemeinschaft (Project No. 413220201).
We acknowledge also support by the European Commission through the N-LIGHT Project within the H2020-MSCA-RISE-2019 call (GA 872196)
and the EIC Pathfinder Project TECHNO-CLS
(Project No. 101046458).
We acknowledge the Frankfurt Center for Scientific Computing (CSC) for providing computer facilities.

\appendix

\section{Estimation of $\Delta \E$ for a positron planar channeling \label{Estimation}}

In the channeling regime the velocity of a projectile is oriented (on
average) along the plane while the interplanar electric field $\bfE$ is
normal to the plane so that $\langle \bfv \rangle \perp \bfE$.
Hence, accounting for the conditions
\begin{eqnarray}
{\bbeta\cdot\bfE \over E} \ll 1,
\qquad
\left(\bbeta\cdot\bfE\right)\bfE \to 0
\label{CaseStudy3:eq.01}
\end{eqnarray}
in  (\ref{Equations:eq.05}) one writes the radiative reaction force
as follows
\begin{eqnarray}
\bfF_{\rm rr}
\approx
-{2 \over 3}  r_0 {\gamma^2 \over mc^2} \Bigl( eE\Bigr)^2 \bbeta
\label{CaseStudy3:eq.02}
\end{eqnarray}

To estimate the energy loss by an ultra-relativistic projectile due to
radiation emission when passing through a crystal of thickness $L$ along
$\bbeta$ ($\beta_z\approx 1$, $\beta_{x,z}\ll 1$) one derives:
\begin{eqnarray}
{\Delta \E \over \E}
\approx 
{1 \over \E}
\int_0^L
\left|F_{{\rm rr},z}\right| \d z
\approx 
{2\E r_0\over 3 (mc^2)^3} \langle e^2 E^2\rangle L
\label{CaseStudy3:eq.04}
\end{eqnarray}
where $\langle e^2 E^2\rangle$ stands for the average value of the squared
force $e\bfE$ which acts on the channeling particle in a crystal.

Assuming the  harmonic approximation for the interplanar potential
\begin{eqnarray}
U(\rho) = {\kappa \rho^2 \over 2},
\qquad
\kappa = {8U_0\over d^2}\,,
\label{CS_3_02:eq.01}
\end{eqnarray}
where $\rho=[-d/2,d/2]$ is the distance from the mid-plane,
$d$ - interplanar distance, $U_0$ is the depth of the well.
The force acting on the projectile is $eE=\kappa \rho$, therefore,
averaging over one period of channeling oscillations
with the amplitude $a$ can be carried out analytically:
\begin{eqnarray}
\langle e^2 E^2\rangle_a
=
{1\over T} \int_{0}^{T} e^2 E^2 \d t
=
\langle e^2 E^2\rangle_{\max}
\left({2a \over d}\right)^2 
\label{CS_3_02:eq.03}
\end{eqnarray}
where
\begin{eqnarray}
\langle e^2 E^2\rangle_{\max}
=
\langle e^2 E^2\rangle_{a=d/2}
=
{8U_0^2\over d^2} 
\label{CS_3_02:eq.03a}
\end{eqnarray}
Averaging over the amplitudes produces
\begin{eqnarray}
\langle e^2 E^2\rangle
=
{2\over d} \int_{0}^{d/2} \langle e^2 E^2\rangle_a \d a
{\kappa^2 d^2\over 24} 
= {\langle e^2 E^2\rangle_{\max} \over 3}
\label{CS_3_02:eq.04}
\end{eqnarray}

\begin{table}[h]
\caption{
Parameters of the (110) interplanar potentials 
(at $T=300$ K)
in diamond (C), Si, Ge crystals:
$d$ - the interplanar distance,
$U_0=U(d/2)$ - the potential well depth,
$\langle e^2 E^2\rangle$ - see Eq. (\ref{CS_3_02:eq.04}),
$\langle e^2 E^2\rangle_{\max}$ - see Eq. (\ref{CS_3_02:eq.03a}).
}
\begin{indented}%
\footnotesize\rm\item[]
\begin{tabular}{@{}ccccccccc}
\br
        &$d$   & $U_0$ & $\langle e^2 E^2\rangle_{\max}$ \\
        &(\AA) & (eV)  &     (GeV$^2$/cm$^2$)            \\
\br
C (110) & 1.26  &  23  &  26.7                           \\
Si(110) & 1.92  &  22  &  10.5                           \\
Ge(110) & 2.00  &  35  &   24.5                          \\
\br
\end{tabular}
\end{indented}
\label{Table.3}
\end{table}

Using the values from Table \ref{Table.3} one derives
the following estimates for the relative radiative losses
of a positron planar channeling in several oriented crystals:
\begin{eqnarray}
{\Delta \E \over \E}
&=
1.4\times10^{-3}\,\E \langle e^2 E^2\rangle L
=
\E\, L
\times
 \left\{
 \begin{array}{ll}
 0.012 & \mbox{diamond(110)}\\
 0.005 & \mbox{Si(110)}\\
 0.011 & \mbox{Ge(110)}
 \end{array}
\right.
\label{CS_3_02:eq.05a}
\end{eqnarray}
On the right-hand side $\E$ and $L$ are measured in GeV and cm,
respectively.

\section{Radiative energy loss in a point Coulomb field: Classical approach
\label{Single}}

In case when an impact parameter $b$ in a collision of an ultra-relativistic
projectile with an atom is much less than the atomic screening radius $R$
one can estimate the radiative energy loss considering the projectile's
motion in the Coulomb field of the nucleus.
In the small angle scattering limit one assumes the uniform motion
along a straight line (the $z$-axis),
$\bfr  \approx \bfb + z \bbeta$.
Substituting the electrostatic force
$\bfF_{\rm em} = Ze^2 (\bfb + z \bbeta)/r^3$ into
Eq. (\ref{Equations:eq.05}) and carrying out the intermediate algebra one
derives the following expression for the relative energy loss
$(\Delta \E/\E)_s$ in a single
close ($b^2 \ll R^2$) collision:
\begin{eqnarray}
\fl
\left(\left.{\Delta \E  \over \E}\right|_{b < R}\right)_s
&=
{1\over \E}
\left|
\int
\bfF_{\rm rr}\cdot \d \bfr
\right|
\approx
{1\over \E}
\left|
\int_{-\infty}^{\infty}
\left(\bfF_{\rm rr}\cdot\bbeta\right)
\d z
\right|
\approx
\gamma {\pi Z^2\over 4}
\left({r_0\over R}\right)^3
\left({R \over b}\right)^3
\label{Yukawa:eq.06}
\end{eqnarray}
Here $r_0=2.818\times 10^{-13}$ is the classical electron radius.
For the estimation purposes,  the Thomas-Fermi radius
can be chosen to characterize the screening
radius $R$:
\begin{eqnarray}
R = \aTF = 0.8853 Z^{-1/3} a_0 
\label{Yukawa:eq.07}
\end{eqnarray}
where $a_0$ stands for the Bohr radius.
For a silicon atom $\aTF = 0.194$ \AA{}.

Within the same approximation, the scattering angle $\theta_s\ll 1$  as a function of the impact parameter $b$ is given by
(see, e.g. \cite{Landau1}):
\begin{eqnarray}
\theta_s
\approx
{1\over \E}
\left|\int_{-\infty}^{\infty} \left(\bfF_{\rm em}\cdot\bbeta\right)\,
\d z\right|
=
{2Z \over \gamma} {r_0\over R}{R \over b}\,.
\label{Classical:eq.02}
\end{eqnarray}

\section*{References}
%

\end{document}